\documentclass[aps,prd,secnumarabic,amssymb, amsmath,nobibnotes,nofootinbib,12pt]{revtex4} 
\usepackage{amsfonts,amsmath,hyperref,url, color}
\usepackage{bm, bbm}
\usepackage{graphicx}
\usepackage{mathtools}
\usepackage{bbold}
\def\N{\mathsf{N}}
\def\A{\mathsf{A}}
\def\K{\mathsf{K}}
\def\SO{\mathsf{SO}}

\usepackage{xcolor}

\usepackage{color}

\usepackage[utf8]{inputenc}
\begin{document}

\title{A group theoretic description of the $\kappa$-Poincar\'e Hopf algebra}

\author{Michele Arzano}
\email{michele.arzano@na.infn.it}
\affiliation{Dipartimento di Fisica ``E. Pancini", Universit\`a di Napoli Federico II, I-80125 Napoli, Italy\\}
\affiliation{INFN, Sezione di Napoli,\\ Complesso Universitario di Monte S. Angelo,\\
Via Cintia Edificio 6, 80126 Napoli, Italy}
\author{Jerzy Kowalski-Glikman}
\email{jerzy.kowalski-glikman@uwr.edu.pl}
\affiliation{University of Wroc\l{}aw, Faculty of Physics and Astronomy \\ Maksa Borna 9, 50-204 Wroc\l{}aw, Poland\\}
\affiliation{National Centre for Nuclear Reserach,\\Pasteura 7, 02-093 Warsaw, Poland}

\begin{abstract}
It is well known in the literature that the momentum space associated to the $\kappa$-Poincar\'e algebra is described by the Lie group $\A\N(3)$. In this letter we show that the full $\kappa$-Poincar\'e Hopf algebra structure can be obtained from rather straightforward group-theoretic manipulations starting from the Iwasawa decomposition of the of the $\SO(1,4)$ group.
\end{abstract}

\maketitle

A key question in the field of quantum gravity is whether there exists a regime in which phenomena induced by the quantum properties of the space-time geometry could possibly be observed. The first proposals for probing the quantum nature of space-time date to about twenty years ago \cite{Amelino-Camelia:1999hpv} and over the years the study of quantum gravity phenomenology has developed into a full fledged research program (see \cite{Addazi:2021xuf} for an up-to-date review). Models of deformed relativistic kinematics based on $\kappa$-deformations of the Poincaré algebra \cite{Lukierski:1991pn,Lukierski:1992dt,Lukierski:1993wx,Lukierski:1993wxa,Majid:1994cy,Kosinski:1994br} (see also \cite{Arzano:2021scz}) have been among the key players in the list of test theories which could lead to potentially observable effects \cite{Amelino-Camelia:2008aez,Amelino-Camelia:2000cpa}. Motivation for considering such models comes from studies of quantum gravity in 2+1 dimensions \cite{Freidel:2005me,Arzano:2013sta,Arzano:2014ppa,Cianfrani:2016ogm} and there are reasons to believe that $\kappa$-deformed relativistic symmetries might be relevant in describing at least some regimes of quantum gravity in the physical 3+1 dimensional case \cite{Amelino-Camelia:2003ezw,Freidel:2003sp}. There is also an intrinsic interest in the study of such models since they provide self-consistent mathematical models of relativistic symmetries and kinematics with a built-in observer independent energy scale with non-trivial structures subject to subtle physical interpretations \cite{Amelino-Camelia:2001rtw,Agostini:2006nc,Arzano:2007ef,Arzano:2007nx,Arzano:2008bt,Arzano:2010jw,Arzano:2014cya,Arzano:2014jfa,Arzano:2017uuh,Arzano:2018gii}. 

Historically, the $\kappa$-Poincar\'e algebra was derived using the tools of the theory of quantum groups as a contraction of the Hopf algebra $\SO_q(3,2)$ \cite{Lukierski:1991pn,Lukierski:1992dt}, with the aim of providing a model which would play a role in Planck-scale physics analogous to that played by the Poincar\'e algebra in quantum field theory. 
A few years later \cite{Majid:1994cy} this algebra was re-derived and brought into the modern form with the help of the general concept of `matched pair of Lie algebras/groups'. In this paper the relation of $\kappa$-Poincar\'e algebra with non-commutative  geometry proposed in \cite{Lukierski:1993wxa} was confirmed. In all these developments advanced techniques of the theory of Hopf algebras were used.
 
{ A notable property of models based on $\kappa$-deformed symmetries is that they are characterized by curved momentum space being the manifold of the Lie group $\A\N(3)$, which geometrically is a submanifold of de Sitter space \cite{Kowalski-Glikman:2002oyi, Kowalski-Glikman:2003qjp}. 
The aim of the present letter is to show that the whole Hopf algebra structure of the $\kappa$-Poincar\'e algebra can be derived from simple group theoretical arguments concerning the action of the Lorentz group $\SO(1,3)$ on $\A\N(3)$ as described from the Iwasawa decomposition of $\SO(1,4)$ of which both $\SO(1,3)$ and $\A\N(3)$ are subgroups. We derive $\kappa$-Poincar\'e Hopf algebra in a language, which is easily understandable to physicists, using only basic group theoretical ingredients. This construction is motivated by, and is an extension of, the notes of S.\ Nowak and one of the present authors \cite{Kowalski-Glikman:2004lyj}. Although the result of our construction is not new, {\it it is the construction itself that is novel}, somewhat unexpected, and will likely turn out to be very useful. This is particularly significant in the context of physical applications of models based on the $\kappa$-Poincar\'e algebra given their relevance for the field of quantum gravity phenomenology (see the recent review \cite{Addazi:2021xuf}).}

The results derived here will be crucially important in a forthcoming paper, in which we propose a resolution of the long standing puzzle of multiparticle states in $\kappa$-deformed quantum field theory. For related investigations see \cite{Sitarz1997},
\cite{Majid:2006xn},
\cite{Gubitosi:2011hgc} and also \cite{Stachura:2015muw}, \cite{Stachura:2018nzh}. All the details of the Hopf-algebraic aspect of our investigations can be found in the monograph \cite{Majid:1996kd}.

{Finally, let us remark that, as it is customary in many physical applications, we are mostly interested here in investigating explicitly the properties of infinitesimal symmetry transformations, although the many properties of finite transformations are also implicit here.}

The starting point of our considerations, as mentioned above, is the Iwasawa decomposition of a Lie algebra and the associted Lie group \cite{Iwasawa} (see also, eg., \cite{Vilenkin}). We will be concerned here with the five-dimensional Lorentz group $\SO(1,4)$. Its Lie algebra $\mathfrak{so}(1,4)$ can be uniquely decomposed into the following direct sum of its subalgebras
\begin{equation}\label{3.18}
    \mathfrak{so}(1,4)=\mathfrak{k}\oplus\mathfrak{a}\oplus\mathfrak{n}
\end{equation}
where $\mathfrak{k}$ denotes the Lorentz algebra $\mathfrak{so}(1,3)$, the algebra  $\mathfrak{a}$ is generated by an element\footnote{In the consideration below we choose to display explicitly the dimensionful scale of deformation $\kappa$, so that all the functions and operators have the canonical physical dimensions. Of course one could use the units in which $\kappa=1$, but then the physical meaning of the resulting formulas would be less transparent.}
 \begin{equation}\label{3.19}
    X^0=-\frac{i}{\kappa}\,\left[\begin{array}{ccc}
    0 &0 & 1 \\
    0 & \mathbb{0} & 0 \\
    1& 0 & 0  \
  \end{array}\right]
\end{equation}
where $\mathbb{0}$ is the $3\times3$ zero-matrix, while the algebra $\mathfrak{n}$ is generated by matrices of the form
\begin{equation}\label{3a1}
  X^i
    =\frac{i}{\kappa}\,\left[ \begin{array}{ccc}
 0 & \epsilon_i & 0 \\
 (\epsilon_i)^T  & \mathbb{0}  & (\epsilon_i)^T \\
    0 & -\epsilon_i & 0
\end{array}\right]
\end{equation}
where $\epsilon_i$ are unit vectors in the direction $i$ ($\epsilon_1 =(1,0,0)$, ect), while $T$ denotes transposition. These matrices can be seen as generators of the product algebra $\mathfrak{an}$ with the following commutators 
\begin{equation}\label{3.7}
    [X^{0},X^{i}]=\frac{i}{\kappa}X^{i}\,, \quad [X^{i},X^{j}]=0\,.
\end{equation}
This algebra, with the generators interpreted as coordinates of a non-commutative generalization of four-dimensional Minkowski space-time, are known in the literature as the $\kappa$-Minkowski non-commutative space-time \cite{Majid:1994cy}.


An arbitrary element of the group $G\in \SO(1,4)$ can be uniquely decomposed into the product of three elements
\begin{equation}\label{Iwagr}
   G= \K\A\N
\end{equation}
 where $\K$ is an element of the Lorentz group $\SO(1,3)$, $\A$ is a group element generated by
${X^0}$ while $\N$ belongs to the group generated by $X^i$. A generic element of the product group $\A\N$ (often denoted by $\A\N(3)$ to stress its dimensionality, in this note for brevity we will use the simpler notation $\A\N$) can be thus written in terms of the matrices \eqref{3.19} and \eqref{3a1}
\begin{align}\label{a1}
    g=e^{ip_iX^i}e^{ip_0X^0}\,.
\end{align}
 The coefficients $p_\mu$ of these exponentials are coordinates on the group manifold which can be seen as functions on the group $\A\N$: $p_\mu = p_\mu(g)$. Using the matrix representation of the generators \eqref{3.19}, \eqref{3a1} the group element \eqref{a1} can be computed to have the form 
 \begin{equation}\label{gcla}
     g=\frac1\kappa\left(
  \begin{array}{ccc}
\tilde P_4&\kappa \mathbf{P}/P_+& P_0\\
\mathbf{P}&\kappa\mathbbm{1}&\mathbf{P}\\
\tilde P_0&-\kappa \mathbf{P}/P_+&P_4\\
  \end{array}
\right)
 \end{equation}
where $\mathbbm{1}$ is the unit $3\times3$ matrix and the non-trivial entries of the matrix are related to the coordinates $p_{\mu}$ via 
\begin{align}\label{classical basis}
 P_{0}&=\kappa\,\sinh\frac{p_{0}}{\kappa} +\frac{\mathbf{p}{}^{2}}{2\kappa}
 e^{{p_{0}}/{\kappa}}\nonumber\\
 P_{i}&= p_{i}e^{{p_{0}}/{\kappa}}\nonumber\\
 P_{4}&=\kappa\,\cosh\frac{p_{0}}{\kappa} -\frac{\mathbf{p}{}^{2}}{2\kappa}
 e^{{p_{0}}/{\kappa}}   \nonumber\\
 P_+ &= P_0 + P_4 =\kappa\, e^{{p_{0}}/{\kappa}}
\end{align}
and 
\begin{align}
\tilde P_0 & =\kappa\, \sinh\frac{p_{0}}{\kappa} -\frac{\mathbf{p}{}^{2}}{2\kappa} e^{{p_{0}}/{\kappa}}\nonumber\\
\tilde P_4 & =\kappa\, \cosh\frac{p_{0}}{\kappa} +\frac{\mathbf{p}{}^{2}}{2\kappa} e^{{p_{0}}/{\kappa}}\,.
\end{align}
It is easily seen that the coordinates $(P_0, \mathbf{P}, P_4)$ are not independent as they satisfy the constraint
\begin{equation}\label{dsde}
-P_0^2 + \mathbf{P}^2 + P_4^2 = \kappa^2    
\end{equation}
which defines a four-dimensional de Sitter space with  cosmological constant equal to $\kappa^{-2}$ embedded in a five-dimensional Minkowski space. The inequality $P_+>0$ restricts us to half of the manifold and thus the momentum space Lie group $\A\N$ is described by half of de Sitter space.

The four-momenta associated to the coordinates $(p_0,\bf{p})$ are the eigenvalues of so called ``bicrossproduct basis" \cite{Majid:1994cy,Borowiec:2009vb} of generators of translations for the $\kappa$-Poincaré algebra $ \mathcal{P}^{B}_{\mu}$ acting on a {\it non-commutative} plane wave \eqref{a1} \cite{Amelino-Camelia:1999jfz}. More precisely, we define a one-particle state to be parametrized by the group element,  $|g\rangle$, then
\begin{equation}
    \mathcal{P}_{\mu}^{B}\, |g\rangle = p_{\mu}(g)\, |g\rangle\,,
\end{equation}
where $p_\mu(g)$ are the coefficients $p_0, p_i$ in \eqref{a1}.

The embedding coordinates $(P_0,\bf{P})$ are instead associated to the so-called ``classical basis" of the $\kappa$-Poincaré algebra \cite{Kosinski:1994br} and whose generators of translations $ \mathcal{P}^{C}_{\mu}$ are such that 
\begin{equation}
    \mathcal{P}_{\mu}^{C}\, |g\rangle = P_{\mu}(g)\, |g\rangle\,,\quad \mathcal{P}_{4}^{C}(g)\, |g\rangle = P_{4}(g)\, |g\rangle\,
\end{equation}
where $P_\mu(g)$ and $P_4(g)$ are the entries in the last column of the matrix \eqref{gcla}.
Therefore, the operators $\mathcal{P}_{\mu}^{C}(g), \mathcal{P}_{4}^{C}(g)$ compute the last column of the matrix \eqref{gcla}. The first four entries of this column are coordinates on momentum space $\A\N(3)$ and the last, as we will see in a moment, is the Lorentz-invariant $\kappa$-Poincar\'e Casimir. We will use this fact frequently in what follows.

As we will soon see, the commutator structure of the generators of the $\kappa$-Poincaré algebra in this basis is just the ordinary Poincaré one. The generators $\mathcal{P}_{\mu}^{B}$ and $\mathcal{P}_{\mu}^{C}$ are directly related to derivatives of the non-commutative differential calculi, see \cite{Sitarz:1994rh,Freidel:2007hk,Rosati:2021sdf,Arzano:2009ci}.

We will now show how the non-abelian nature of the momentum space leads to a non-abelian composition of four-momenta. To this end, let us consider elements of $\SO(1,4)$ belonging to the $\A\N$ subgroup (for which the first factor in \eqref{Iwagr} is the identity $\K =\mathbbm{1}$, of the form \eqref{a1}. Using the commutation relations \eqref{3.7} or the explicit matrix form of $\A\N$ elements one can show that 
\begin{equation}\label{3.37}
    e^{ip_{i}X^{i}}e^{ip_{0}X^{0}}=e^{ip_{0}X^{0}}e^{ip_{i}e^{p_{0}/\kappa}X^{i}}
\end{equation}
and thus for the product of two group elements we have
\begin{align}
gh=    e^{ip_{i}X^{i}}e^{ip_{0}X^{0}}e^{iq_{i}X^{i}}e^{iq_{0}X^{0}}=e^{ i\left(p_{i}+q_{i}\,e^{-p_{0}/\kappa}\right)X^{i}}e^{i(p_{0}+q_{0})X^0}\label{3.36}
\end{align}
which determines the non-abelian addition of four-momenta
\begin{equation}\label{oplus}
p \oplus q = \left(p_{0}+q_{0},\mathbf{p}+\mathbf{q}e^{-p_{0}/\kappa}\right)\,.
\end{equation}
 These deformed composition rules for four-momenta reflect a non trivial action of translation generators on tensor product representations \cite{Arzano:2021scz}. 
 
Now let us define the map $\Delta(P^{B}_{\mu})$, known in Hopf algebra language as the {\it coproduct}, which defines the action of translations (momenta) on the a two particles state, being an element of the tensor product of two one-particle Hilbert spaces
\begin{align}\label{coproddef}
     \Delta(\mathcal{P}_\mu^{B}) \, |g\rangle \otimes |h\rangle \equiv p_\mu(gh)\, |g\rangle \otimes |h\rangle  = (p \oplus q )_\mu \, |g\rangle \otimes |h\rangle 
\end{align}
which with the help of \eqref{oplus} can be solved to give
\begin{align}
 \Delta(\mathcal{P}^{B}_{i})=\mathcal{P}^{B}_i\otimes \mathbbm{1} +e^{-{\mathcal{P}^{B}_0}/{\kappa}}\otimes P^{B}_i\,,\quad
  \Delta(\mathcal{P}^{B}_{0}) = \mathcal{P}^{B}_{0}\otimes \mathbbm{1} +  \mathbbm{1} \otimes \mathcal{P}^{B}_{0}.
\end{align}
This is the standard way one derives the co-product for the deformed translation generators of the $\kappa$-Poincaré algebra and can be found in various works in the literature. 

Returning to the $\A\N(3)$ group, for the inverse group element one has 
\begin{align}
 \left(e^{ip_{i}
X^{i}}e^{ip_{0}X^{0}}\right)^{-1}=e^{-ip_{i}e^{p_{0}/\kappa}X^i}e^{-ip_{0}X^{0}} \label{3.38} 
\end{align}
from which we can define the following map on four-momenta 
\begin{equation}\label{invbicr}
\ominus p \equiv \left(-p_{0}, -\mathbf{p}\, e^{p_{0}/\kappa}\right)    
\end{equation}
which is the non-abelian analogue of the minus operation on usual four-vectors ensuring that $p\oplus (\ominus(p)) =(\ominus(p)) \oplus p= 0$.

The inverse group element relation \eqref{invbicr} make it possible to define another important concept of the {\it antipode} denoted by $S$ and defined by the condition
\begin{align}
\label{antipo}
   S( \mathcal{P}_{\mu}^{B})) |g\rangle = \mathcal{P}^{B}_{\mu} |g^{-1}\rangle 
\end{align}
which can be solved to give 
\begin{align}\label{antipP}
    S(\mathcal{P}^{B}_{0})=-\mathcal{P}^{B}_{0}\,, \quad S(\mathcal{P}^{B}_{i})=-\mathcal{P}^{B}_{i}e^{\mathcal{P}^{B}_{0}/\kappa}\,.
\end{align}

Let us pause here to explain the physical meaning of the notions introduced above. As we saw, the coproduct tells us how to consistently define the action of a symmetry generator (in the case above, the momentum) on multiparticle states. The (co)-associativity of the coproduct guarantees that the action on two particles state can be generalized to the action on all multiparticle states. The antipode instead provides an information about the action of the inverse symmetry generators. In the case of generators different from momenta (mathematically, functions on the group) to compute the antipode one has to use a more general construction, which we will explain when discussing Lorentz transformations.

Let us now consider an alternative, more straightforward derivation of the coproduct which makes use of the product of two momentum matrices \eqref{gcla}
 \begin{equation}
     g h =\frac1{\kappa^2}\, \left(
  \begin{array}{ccc}
\tilde P_4&\kappa  \mathbf{P}/P_+& P_0\\
\mathbf{P}&\kappa \mathbbm{1}&\mathbf{P}\\
\tilde P_0&-\kappa  \mathbf{P}/P_+&P_4\\
  \end{array}
\right)
\left(
  \begin{array}{ccc}
\tilde Q_4&\kappa   \mathbf{Q}/Q_+& Q_0\\
\mathbf{Q}&\kappa \mathbbm{1}&\mathbf{Q}\\
\tilde Q_0&-\kappa \mathbf{Q}/Q_+&Q_4\\
  \end{array}
\right)\,.
 \end{equation}
Remembering that $P_\mu(gh)$ reads
off the first four components of the fifth column of the resulting matrix on the right hand side we get the following non-abelian composition of momenta $ P_{\mu}(g) \oplus P_{\mu}(h) \equiv P_{\mu}(g h)$:
\begin{align}
    P_0\oplus Q_0 & =\frac1\kappa\, \tilde{P}_4\, Q_0 + \frac{\mathbf{P}\cdot \mathbf{Q}}{P_+} +\frac1\kappa\,  P_0\, Q_4= \frac\kappa{P_+} Q_0 + \frac{\mathbf{P}\cdot \mathbf{Q}}{P_+} +\frac1\kappa\,  P_0\, Q_+ \\
    \mathbf{P}\oplus \mathbf{Q} & = \frac1\kappa\, \mathbf{P}\, Q_+ + \mathbf{Q}
\end{align}
Using the defining property
\begin{align}\label{coproddefC}
     \Delta(\mathcal{P}_\mu^{C}) \, |g\rangle \otimes |h\rangle \equiv P_\mu(gh)\, |g\rangle \otimes |h\rangle  = (P \oplus Q )_\mu \, |g\rangle \otimes |h\rangle 
\end{align}
we thus find
\begin{align}
\Delta(\mathcal{P}^{C}_{0}) &= \kappa \, \left(\mathcal{P}^{C}_{+}\right)^{-1}\otimes \mathcal{Q}^{C}_{0} + \sum_i \left(\mathcal{P}^{C}_{+}\right)^{-1} \mathcal{P}^{C}_{i}\otimes \mathcal{Q}^{C}_{i} +\frac1\kappa \, \mathcal{P}^{C}_{0}\otimes \mathcal{Q}^{C}_{+} \\
\Delta(\mathcal{P}^{C}_{i}) &=  \frac1\kappa \, \mathcal{P}^{C}_{i}\otimes \mathcal{Q}^{C}_{+} + \mathcal{Q}^{C}_{i}\otimes \mathbbm{1}\,.
\end{align}
From the condition $P_{\mu}\oplus(\ominus P_{\mu})= 0$ one can also easily derive
\begin{align}
    \ominus P_0 & = -P_{0}+\mathbf{P}^{2}/P_{+}\\
    \ominus \mathbf{P} & =  -\kappa \mathbf{P}/P_+\,. 
\end{align}
so that
\begin{align}
    S(\mathcal{P}^{C})_{0} & = -\mathcal{P}^{C}_{0}+(\mathcal{P}^{C}_{i})^2 \left(\mathcal{P}^{C}_{+}\right)^{-1}\\
  S(\mathcal{P}^{C})_{i}   & =  -\kappa \mathcal{P}^{C}_{i} \left(\mathcal{P}^{C}_{+}\right)^{-1}\,. 
\end{align}

Let us now turn to unveiling the Hopf algebra structure of the Lorentz sector, i.e.\ the coproduct and antipodes for the generators of boosts and rotations. To this end we must first spell out the action of the Lorentz group on the group manifold momentum space $\A\N$. This can be done by noticing that the Iwasawa decomposition guarantees that given the Lorentz group element $K$ and $g$ belonging to $\A\N$ group, there are unique $K'_{g}$ and $g'$ satisfying the equality
\begin{equation}\label{a2}
   K\, g = g'\, K'_{g}\,. 
\end{equation}
Since $K$ is a Lorentz transformation we \textit{define} $g'$ to be the Lorentz-transformed $g$, and we  write 
\begin{equation}\label{a3}
  K\, e^{ip_iX^i}e^{ip_0X^0} \, K'_{g}{}^{-1}= e^{ip'_iX^i}e^{ip'_0X^0} 
\end{equation}
where the four-momenta $(p_0', p_i')$ are defined to be resulting from the Lorentz action on $(p_0, p_i)$. For infinitesimal Lorentz transformation we assume that
\begin{equation}\label{linLor}
    K \approx \mathbbm{1}+i\xi^a \mathfrak{K}_a\,,\quad K'_{g} \approx \mathbbm{1}+i\xi^a \, h_a^b(g) \mathfrak{K}_b
\end{equation}
where $\mathfrak{K}_a$ are the generators of Lorentz algebra $\mathfrak{so}(1,3)$, $\mathfrak{K}_a=(N_i,M_i)$ with $N_i$, $M_i$ denoting the generators of boost and rotations, and $h_a^b(g) $ is a momentum-dependent matrix that we will explicitly derive below. As we will see this function is responsible for the effect of `back-reaction' of momenta on Lorentz transformations on products of momentum group elements, making the effective infinitesimal parameter of Lorentz transformation momentum dependent, as noticed in \cite{Majid:2006xn} (see also \cite{Gubitosi:2011hgc}). We will return to this point below.

In order to derive the explicit form of infinitesimal Lorentz transformation on four-momenta we focus on the the parametrization of the group element \eqref{gcla} in terms of coordinates $P_\mu$. For definiteness, and without loss of generality we consider  a boost in the 1-direction. In matrix form equation \eqref{a2} can be written explicitly as 
\begin{align}
    \left(
  \begin{array}{ccccc}
 1&\xi &0&0&0\\
\xi &1&0&0&0\\
0&0&1&0&0\\
0&0&0&1&0\\
0&0&0&0&1\\
  \end{array}
\right)\left(
  \begin{array}{ccccc}
\tilde P_4& \kappa P_1/P_+&\kappa  P_2/P_+&\kappa P_3/P_+&P_0\\
P_1&\kappa &0&0&P_1\\
P_2&0&\kappa &0&P_2\\
P_3&0&0&\kappa &P_3\\
\tilde P_0&-\kappa  P_1/P_+& -\kappa P_2/P_+&-\kappa P_3/P_+&P_4\\
  \end{array}
\right)\nonumber\\
=\left(
  \begin{array}{ccccc}
\tilde P'_4&\kappa  (P_1/P_+)'&\kappa  (P_2/P_+)'&\kappa (P_3/P_+)'&P'_0\\
P'_1&\kappa &0&0&P'_1\\
P'_2&0&\kappa &0&P'_2\\
P'_3&0&0&\kappa &P'_3\\
\tilde P_0&-\kappa  (P_1/P_+)'&-\kappa  (P_2/P_+)'&-\kappa (P_3/P_+)'&P_4\\
  \end{array}
\right)
\left(
  \begin{array}{ccccc}
 1&\bar\xi_1 &\bar\xi_2&\bar\xi_3&0\\
\bar\xi_1 &1&\bar\rho_3&-\bar\rho_2&0\\
\bar\xi_2&-\bar\rho_3&1&\bar\rho_1&0\\
\bar\xi_3&\bar\rho_2&-\bar\rho_1&1&0\\
0&0&0&0&1\\
  \end{array}
\right)\label{matrixLorentz}
\end{align}
In the last matrix representing $\mathbbm{1}+i\xi^a h_a^{b}(p)\mathfrak{K}_b$ the entries $\bar\xi_i$ and $\bar\rho_i$ are, respectively, the parameters of the infinitesimal boost and rotation, which by force of Iwasawa decomposition are uniquely defined. As we pointed out above, the embedding coordinate functions $P_{\mu}(g)$ simply read the first four entries of the fifth column of the matrix \eqref{gcla}. We thus have the key property 
\begin{equation}\label{kprop}
  P_{\mu}(g') = P_{\mu}(K\, g \,K_{g}^{'-1}) = P_{\mu}(K\, g)
\end{equation}
which simply tells us that infinitesimally $P'_\mu=P_\mu+ \delta P_\mu$ with
\begin{equation}\label{lnzP}
    P_0' = P_0 + \xi P_1\,,\quad P_1' = P_1 + \xi P_0\,,\quad P_2'=P_2\,,\quad P_3'=P_3\,,\quad P_4'=P_4
\end{equation}
and therefore $P_\mu$ {\it transforms as a ordinary Lorentz vector}. We thus have that the commutators between the generators of boosts and the generators of translations of the classical basis of the $\kappa$-Poincaré algebra are just the undeformed ones
\begin{equation}
    \left[\mathcal{N}_{i}, \mathcal{P}^{C}_{j}\right] = i\,  \delta_{ij} \mathcal{P}^{C}_{0},\,\,\, \left[\mathcal{N}_{i},\mathcal{P}^{C}_{0}\right] = i\
\mathcal{P}^{C}_{i}\,.
\end{equation}
 Although in what follows we will only need the infinitesimal transformation, it is easy to see that the structure of the matrices in \eqref{matrixLorentz} remains the same in the case of finite transformations so that, for the final boost along the first axis we obtain
\begin{align}\label{lnzPf}
    P_0' =\cosh\xi P_0 + \sinh\xi P_1\,,\quad P_1' =\cosh\xi P_1 +\sinh \xi P_0\,,\quad P_2'=P_2\,,\quad P_3'=P_3\,,\quad P_4'=P_4
\end{align}

Using \eqref{classical basis} we easily derive the boost action on the bicrossproduct coordinates $p_\mu$ which determine the following non-linear commutators between generators of boosts and bicrossproduct generators of translations 
\begin{equation}
       \left[\mathcal{N}_{i}, \mathcal{P}^{B}_{j}\right] = i\,  \delta_{ij}
 \left( {\frac\kappa2 } \left(
 1 -e^{-2\mathcal{P}^{B}_{0}/\kappa}
\right) + \frac{(\mathcal{P}^{B}_i)^2}{2\kappa}  \right) - i\,
\frac1\kappa\, \mathcal{P}^{B}_{i} \mathcal{P}^{B}_{j} ,\,\,\, \left[\mathcal{N}_{i},\mathcal{P}^{B}_{0}\right] = i\
\mathcal{P}^{B}_{i}\,.
\end{equation}
Repeating the steps leading to \eqref{lnzP} one finds that infinitesimal rotations act in the ordinary way on the four-momenta $P_{\mu}$ and through the coordinate transformation \eqref{classical basis} it is immediate to see that also the bicrossproduct momenta $p_{\mu}$ transform in the standard way and thus we have the commutators
\begin{equation}
    [\mathcal{M}_i, \mathcal{P}^{B}_j] = i\, \epsilon_{ijk} \mathcal{P}^{B}_k, \quad [\mathcal{M}_i, \mathcal{P}^{B}_{0}] =0\,.
\end{equation}

From \eqref{lnzP} we can also see that $P_4$ is a Lorentz scalar and thus it must be related to the Casimir opertator. Indeed from this observation and from \eqref{dsde} we see that 
\begin{equation}
    P^2_0-\mathbf{P}^2 = P_4^2-\kappa^{2}
\end{equation}
is an invariant and indeed, since in the classical basis the algebra remains invariant, we have that the classical translation generators can be combined in the standard Casimir operator
\begin{equation}
    \mathcal{C} = (\mathcal{P}^{C}_{0})^2 - (\vec{\mathcal{P}}^{C})^2\,.
\end{equation}

With simple algebraic manipulations we have thus derived the important result that in the bicrossproduct basis boosts act non-linearly on four-momenta (while rotations transform momenta in the ordinary way) while momenta defined by embedding coordinates on the $\A\N$ group manifold transform under the ordinary action of boosts and rotations. As we anticipated, the $\kappa$-Poincaré algebra in the classical basis associated to the translation generators $\mathcal{P}^C_{\mu}$ closes an ordinary Poincaré algebra; the non-abelian nature of four-momenta affects only the co-products and antipodes of the algebra generators. In the remainder of this letter we show how to derive co-products and antipodes for the generators of boosts and rotations using the tools introduced so far.

Before continuing let us  determine the matrix $h_a^{b}(g)$ in \eqref{linLor}. To this end we now consider the fourth columns of the matrices on both sides of \eqref{matrixLorentz}, 
\begin{align}
    \left(
  \begin{array}{c}
 \kappa  P_3/P_+\\ \xi \kappa P_3/P_+\\ 0\\ \kappa \\ -\kappa P_3/P_+
  \end{array}
\right)  =
    \left(
  \begin{array}{c} 
  \bar\xi_3\, \tilde P'_4 - \bar\rho_2\kappa  (P_1/P_+)' + \bar\rho_1\kappa  (P_2/P_+)' + \kappa (P_3/P_+)'\\ \bar\xi_3 P_1' -\bar\rho_2\kappa \\\bar\xi_3 P_2' +\bar\rho_1\kappa \\\bar\xi_3 P_3' +\kappa \\
   \bar\xi_3\, \tilde P'_0 + \bar\rho_2 \kappa (P_1/P_+)' - \bar\rho_1 \kappa (P_2/P_+)' -\kappa  (P_3/P_+)'
      \end{array}
    \right)
\end{align}
from which we get the conditions $\bar\xi_3=0$ and $\bar\rho_1=0$. Analogously, form the third column we derive $\bar\xi_2=0$, so that the only remaining boost parameter is $\bar\xi_1\equiv\bar\xi$. Further, comparing the third and the fourth columns on both sides we find that
\begin{align}\label{barrho}
    \bar\rho_3 =\xi\, \frac{P_2}{P_+}\,,\quad \bar\rho_2 = -\xi\, \frac{P_3}{P_+}
\end{align}
Finally, we compare the second columns, and from the second rows we find the equation
$$
\xi\kappa  \frac{P_1}{P_+}+\kappa  = \bar\xi P_1'+\kappa 
$$
Since $P_1' = P_1 +\xi P_0$, working in the linear order we derive
\begin{align}\label{barxi}
\bar\xi = \frac\kappa {P_+}\,  \xi
\end{align}\\
From \eqref{barrho} and \eqref{barxi} we deduce the form of the matrix $i\xi^ah_a^b(g)\mathfrak{K}_b$ for $\xi^a =(\xi, 0,0)$ 
\begin{align}
i\xi^ah_a^b(g)\mathfrak{K}_b=\frac{1}{P_+}\,\xi\left(
  \begin{array}{ccccc}
 0&\kappa  &0&0&0\\
\kappa  &0&P_2&P_3&0\\
0&-P_2&0&0&0\\
0&-P_2&0&0&0\\
0&0&0&0&0\\
  \end{array}
\right)\label{matrixh}
\end{align}
This can be easily generalized to the case of a boost in an arbitrary direction, with infinitesimal parameters $\xi^i$. The matrix in this case will consist of a combination of the infinitesimal boost  $\bar\xi^i\, N_i$ and infinitesimal rotation $\bar\rho^i\, M_i$ with
\begin{align}\label{hab1}
  \bar  \xi^i= \xi^i \frac{\kappa }{P_+} = \xi^i\, e^{-p_0/\kappa}
\end{align}
and 
\begin{align}\label{hab2}
  \bar  \rho^i= \epsilon_{j}{}^{ki}\,\xi^j\, \frac{P_k}{P_+} = \frac1\kappa\, \epsilon_{j}{}^{ki}\,\xi^j\, p_k\,.
\end{align}
Notice that if instead of boost we considered an infinitesimal rotation on the left hand side of \eqref{matrixLorentz} we would find the same rotation on the right hand side, which shows that the matrix $h_i^b(g)$ for rotations is the identity.

Let us now turn to the co-product of Lorentz generators. To this end, we look at the action of the Lorentz group on the product of two $\A\N$ group elements $g$ and $h$. Using the Iwasawa decomposition again we have
\begin{align}
    K gh = (gh)'K'_{gh}
\end{align}
which we can rewrite as
\begin{align}\label{s20}
    (gh)' =  K g K_g^{'-1} \, K_g\, h K_{gh}^{'-1} = g' \, K'_g\, h K_{gh}^{'-1} 
\end{align}
and from which it is clear that $(gh)'\neq g'\, h'$, i.e., in physical terms, that the Lorentz group action on momentum space is not Leibnizian. Using the property \eqref{kprop} we have
\begin{equation}\label{kpropcp}
    P_{\mu}((gh)')= P_{\mu}( g' \, K'_g\, h K_{gh}^{'-1}) 
    = P_{\mu}(g' K'_g\, h)
\end{equation}
where the last equality comes from the property \eqref{kprop}.
We see that, unlike the case of Lorentz transformations of single group elements, now the matrix $K'_g$ plays a role in determining how the Lorentz boosted momentum obtained by a product of two group elements can be expressed in terms of the transformations of the individual factors. 

In order to determine the coproduct for boost generator let us again consider an infinitesimal boost matrix $K = \mathbbm{1}+ \xi {N}_1 $ and write equation \eqref{s20} in the matrix form
\begin{align}\label{ghprimematrix}
 &  (gh)' =   g' \, K'_g\, h K_{gh}^{'-1} \nonumber\\
& = \frac1{\kappa^2}\left(
  \begin{array}{ccc}
\tilde P'_4&\kappa \left(\mathbf{P}/P_+\right)'& P'_0\\
\mathbf{P}'&\kappa\mathbbm{1}&\mathbf{P}'\\
\tilde P'_0&-\kappa\left( \mathbf{P}/P_+\right)'&P'_4\\
  \end{array}
\right) \, \left(
  \begin{array}{ccccc}
 1&\frac\kappa{P_+}\xi &0&0&0\\
\frac\kappa{P_+}\xi &1&\frac{P_2}{P_+}\xi&\frac{P_3}{P_+}\xi&0\\
0&-\frac{P_2}{P_+}\xi&1&0&0\\
0&-\frac{P_3}{P_+}\xi&0&1&0\\
0&0&0&0&1\\
  \end{array}
\right) \left(
  \begin{array}{ccc}
\tilde Q_4&\kappa \mathbf{Q}/Q_+& Q_0\\
\mathbf{Q}&\kappa\mathbbm{1}&\mathbf{Q}\\
\tilde Q_0&-\kappa \mathbf{Q}/Q_+&Q_4\\
  \end{array}
\right)\, K_{gh}^{'-1}
\end{align}
Keeping only the terms up to the linear order in $\xi$ this equation can be rewritten as
\begin{align}\label{ghprime1}
    (gh)' = \left[gh +\xi\left(\left(N_1 g \right)\left(h\right) + \left(\frac{\kappa}{P_+} g \right)\left(N_1 h\right)+\epsilon_{1jk}\left(\frac{P_j}{P_+} g \right)\left(M_k h\right)    \right)\right]K_{gh}^{'-1}
\end{align}
As we know, the matrix $K_{gh}^{'-1}$ does not influence the momenta, so its explicit form is not relevant. The formula \eqref{ghprime1} can be easily generalized to the case of $K = \mathbbm{1}+ \xi^i {N}_i $.
From \eqref{ghprime1} we can immediately read off the following expression for the co-product of boosts in the classical basis
\begin{equation}\label{coproN}
\Delta^{C} \mathcal{N}_i = \mathcal{N}_i \otimes \mathbbm{1} + \frac\kappa{\mathcal{P}^{C}_+} \otimes \mathcal{N}_i + \epsilon_{ij}{}^k\, \frac{\mathcal{P}^{C}_j}{\mathcal{P}^{C}_+} \otimes \mathcal{M}_k\,. 
\end{equation}
Equation \eqref{coproN} can be rewritten in the bicroproduct basis
\begin{equation}
    \Delta^{B} \mathcal{N}_i = \mathcal{N}_i\otimes \mathbbm{1} +e^{-\mathcal{P}^{B}_0/\kappa}\otimes \mathcal{N}_i+\frac1\kappa\, \epsilon_{ijk}\, \, \mathcal{P}^{B}_j\otimes \mathcal{M}_k\,.
\end{equation}

One can repeat the same procedure for infinitesimal rotations. In this case however, as noticed above, the matrix $h^a_a(g)$ reduces to the identity matrix and thus the generators of rotations act following the the usual Leibniz rule i.e. the have a trivial co-product
\begin{equation}
    \Delta \mathcal{M}_i = \mathcal{M}_i\otimes \mathbbm{1} +\mathbbm{1}\otimes \mathcal{M}_i\,.
\end{equation}

Finally, let us consider the antipode of Lorentz generators. While the co-product determines the way the generators of Lorentz transformations act on the product to two $\A\N$ group elements, the antipode tells what the {\it inverse} Lorentz generator is. Consider \eqref{ghprime1} for $g\mapsto g^{-1}$, $h\mapsto g$
\begin{align}\label{ghprimegg}
 \mathbbm{1}=   (g^{-1}g)' = \left[\mathbbm{1} +\xi^i\left(\left(N_i g^{-1} \right)\left(g\right) + \left(\frac{P_+}{\kappa} g^{-1} \right)\left(N_i g\right)-\epsilon_{ijk}\left(P_j g^{-1} \right)\left(M_k g\right)    \right)\right]K_{g^{-1}g}^{'-1}
\end{align} 
and we see that the term proportional to $\xi^i$ must vanish.

Let us define the antipode of  Lorentz generator by the formula
\begin{align}
    N_i  g^{-1} &\equiv  (g^{-1})S(N_1)\,,\quad
    M_k  g^{-1} \equiv  (g^{-1})S(M_k)
\end{align}
substituting this to \eqref{ghprimegg} we get
\begin{align}\label{antipode1}
    S(N_i) +\frac{P_+}{\kappa} N_i - \epsilon_{ijk}\,{P_j} M_k   =0
\end{align}
Repeating the same procedure for the $\mathbbm{1}= g^{-1} g$ we derive
\begin{align}\label{antipode2}
 N_i  + \frac{\kappa}{P_+} S(N_i) +\epsilon_{ijk}\,\frac{P_j}{P_+} S(M_k)   =0 
\end{align}
Solving  \eqref{antipode1} we find expression for $S(N)$ and substituting the solution to \label{antipode2} we derive $S(M)$. In terms of $\mathcal{N}_i$ and $\mathcal{M}_i$ we have
\begin{align}
    S^{C}(\mathcal{N}_i) = 
    -\frac{\mathcal{P}^{C}_+}{\kappa} \mathcal{N}_i +\frac1\kappa \epsilon_{ijk} \mathcal{P}^{C}_j \mathcal{M}_k
\end{align}
and for rotations
\begin{align}
    S^{C}(\mathcal{M}_i) = - \mathcal{M}_i\,.
\end{align}
The corresponding expressions in the bicrossproduct basis are
\begin{align}
    S^{B}(\mathcal{N}_i) & = -e^{\mathcal{P}^{B}_0/\kappa} \left(\mathcal{N}_i -\frac1\kappa\, \epsilon_{ijk} \mathcal{P}^{B}_j \mathcal{M}_k\right)\\
    S^{B}(\mathcal{M}_i) & = - \mathcal{M}_i\,.
\end{align}

It should be stressed that it is the matrix $h^a_b(g)$ \eqref{matrixh} that captures the essence of all nontrivial structures associated with coproducts and antipodes of the $\kappa$-Poincar\'e algebra. It is also responsible for the emergence of what was called in \cite{Majid:2006xn} and \cite{Gubitosi:2011hgc} the `back-reaction of the momenta on the Lorentz sector'. Indeed,  the first factor $e^{-p_0/\kappa} N_i$ in \eqref{hab1} is the infinitesimal counterpart of the back-reaction found in these papers. 

To conclude, in this note we showed that the whole of the Hopf structure of $\kappa$-Poincar\'e algebra can be straightforwardly deduced from the theory of Iwasawa decomposition of the five-dimensional Lorentz group $\SO(1.4)$. We hope that this construction will help, among others, to to understand better the relations between $\kappa$-deformation and quantum gravity in  four and three dimensions, generalizing the results of \cite{Kowalski-Glikman:2008fix} and \cite{Trzesniewski:2017sqa}.

\section*{Acknowledgment} 

This paper is inspired by unpublished notes by Sebastian Nowak and JKG \cite{Kowalski-Glikman:2004lyj}. We thank Andrea Bevilacqua and Giacomo Rosati for their helpful comments on the early version of the manuscript. For JKG, this work was supported by funds provided by the National Science Center, project number 2019/33/B/ST2/00050. This work contributes to the European Union COST Action CA18108 {\it Quantum gravity phenomenology in the multi-messenger approach.}

\end{document}